\documentclass[useAMS,usenatbib,letterpaper,onecolumn]{mn2e}
\usepackage{geometry }

\title[Polarization transfer in pulsar magnetosphere]
{Polarization transfer in pulsar magnetosphere}
\author[S. A. Petrova]{S. A. Petrova \thanks{E-mail:
petrova@ira.kharkov.ua}\\ Institute of Radio Astronomy, 4,
Chervonopraporna Str., Kharkov 61002, Ukraine}

\begin{document}

\date{Accepted.... Received ....; in original form .....}


\maketitle


\begin{abstract}
Propagation of radio waves in the ultrarelativistic magnetized
electron-positron plasma of pulsar magnetosphere is considered.
Polarization state of the original natural waves is found to vary
markedly on account of the wave mode coupling and cyclotron
absorption. The change is most pronounced when the regions of mode
coupling and cyclotron resonance approximately coincide. In cases
when the wave mode coupling occurs above and below the resonance
region, the resultant polarization appears essentially distinct.
The main result of the paper is that in the former case the
polarization modes become non-orthogonal. The analytical treatment
of the equations of polarization transfer is accompanied by the
numerical calculations. The observational consequences of
polarization evolution in pulsar plasma are discussed as well.
\end{abstract}

\begin{keywords}
waves --- plasmas --- polarization --- pulsars: general
\end{keywords}


\section{Introduction}
\subsection{Empirical model of pulsar polarization}
The radio emission observed from pulsars is typically
characterized by a high percentage of linear polarization. Within
the framework of a well-known rotating vector model \citep{RC69},
the orientation of pulsar polarization reflects the magnetic field
geometry in the emission region and therefore the position angle
(PA) changes monotonically as the pulsar beam rotates with respect
to an observer. The characteristic {\it S}-shaped swing of PA
across the pulse is indeed observed in a number of pulsars. In
addition, PA may show abrupt jumps by approximately $90^\circ$
\citep[e.g.][]{M75}, testifying to the presence of the two
orthogonally polarized modes (OPMs). The early studies of this
phenomenon have revealed that for each of the OPMs PA roughly
follows the predictions of the rotating vector model \citep*{B76}
and mode changing is a stochastic process \citep*{C78}. Further on
the OPMs have been recognized as a fundamental feature of pulsar
radio emission \citep{BR80,S84a,S84b}. A comprehensive analysis of
the observational data has proved an idea of superposed OPMs: At
any pulse longitude the radio emission is believed to present an
incoherent mixture of the two OPMs, whose intensities vary
randomly from pulse to pulse \citep{Mc98,Mc00}.

The plasma of pulsar magnetosphere does allow two types of
non-damping natural waves, the ordinary and extraordinary ones.
They propagate in a superstrong magnetic field, generally at not a
small angle to the field lines, and, correspondingly, are linearly
polarized in orthogonal directions. The electric vector of the
ordinary wave lies in the same plane as the wavevector and the
ambient magnetic field, while the extraordinary wave is polarized
perpendicularly to this plane. The origin of the two types of
natural waves is attributed either to the two distinct radio
emission mechanisms \citep[e.g.][]{Mc97} or to the partial
conversion of the ordinary mode into the extraordinary one
\citep{P01}. A recently discovered anticorrelation of the OPM
intensities \citep{ES04} favours the latter scenario.

It should be noted, however, that direct identification of the
observed superposed OPMs with the natural modes of pulsar plasma
faces serious difficulties. First of all, some circular
polarization is always present in pulsar radiation. Usually it is
much lower than its linear counterpart but not negligible. It has
been noticed that the sign of circular polarization is well
correlated with PA \citep{C78}, that is the two OPMs have circular
polarization of opposite signs and can be regarded as purely
orthogonal elliptical modes. As for the theoretical
interpretation, immediate switching to the case of elliptical
natural waves \citep*{Mel77,Mel79,Kunzl98,Mel04a,Mel04b} seems
problematic. The ellipticity of the natural waves may result from
the gyrotropy of pulsar plasma (caused by difference in the
distributions of electrons and positrons), if only the waves
propagate quasi-longitudinally with respect to the magnetic field.
Although the plasma gyrotropy is very probable, the regime of
quasi-longitudinal propagation is not characteristic of pulsar
magnetosphere. It can be the case only at some specific locations
(most likely, close to the magnetic axis, where the divergence of
magnetic field lines is less significant) and cannot account for
the elliptically polarized natural waves observed throughout the
pulse and over a wide frequency range.

Recent thorough studies of the single-pulse data have introduced
further complications into the picture of pulsar polarization. It
has been found that the observed fluctuations of the Stokes
parameters cannot be explained solely by the pulse-to-pulse
variation of the OPM intensities \citep{Mc04,ES04}. To account for
the observations it has been suggested to complement the OPMs with
a randomly polarized component of unknown nature \citep{Mc04}.
Alternatively, the same results can be interpreted as a
consequence of pulse-to-pulse jitter of both the ellipticity and
PA \citep*{KarIII,ES04}. However, even this generalized picture
based on the OPMs with the randomly varying vector of the Stokes
parameters appears incomplete. In some cases, the modes are
clearly non-orthogonal \citep{KarIII,Mc04,ES04}. It should be
mentioned that the non-orthogonality of polarization modes
manifests itself not only in PA, but also in the circular
polarization. In particular, the same PA may be accompanied by the
circular polarization of any sense, the correlation between PA and
$V$ being less perfect at higher frequencies \citep{KarI,KarIII}.

Diverse and complicated behaviour of the single-pulse polarization
as well as its strong frequency dependence motivate our study of
the propagation effects in pulsar magnetosphere.

\subsection{Wave mode coupling in pulsar magnetosphere}
Pulsar radio emission is believed to be generated deep inside the
tube of open magnetic lines. Then it should propagate through the
flow of an ultrarelativistic highly magnetized electron-positron
plasma. As a result of propagation effects, polarization of the
radio waves may evolve significantly. In the vicinity of the
emission region, the ordinary and extraordinary waves are linearly
polarized in orthogonal directions and the plasma number density
is so large that the geometrical optics regime holds: the natural
waves propagate independently, with the electric vectors being
adjusted to the orientation of the ambient magnetic field. As the
plasma number density decreases with distance from the neutron
star, the difference in the refractive indices of the waves
decreases as well, and finally the scale length for beating
between the modes becomes comparable to the scale length for
change in the plasma parameters. Then the polarization planes of
the waves have no time to follow the local magnetic field
direction, geometrical optics approximation is broken and wave
mode coupling starts. Typically this occurs in the outer
magnetosphere, at distances of a few tenth of the light cylinder
radius. In the region of wave mode coupling, each of the incident
natural waves becomes a coherent sum of both natural waves
peculiar to the ambient plasma, with the amplitude ratio and phase
difference varying along the trajectory. Correspondingly, the
ellipticity of the waves increases with distance and the major
axis of polarization ellipse is monotonically shifted, so that it
no longer reflects the orientation of the ambient magnetic field.
Further on, as the plasma density decreases considerably, the
waves decouple from the plasma and propagate just as in vacuum,
preserving their elliptical polarization. Therefore the process of
wave mode coupling is usually called polarization-limiting effect.

This effect has long been used to explain the origin of circular
polarization in pulsar radio emission \citep{CR79,RR90,LP99}. The
numerical tracings of polarization evolution in pulsar plasma have
demonstrated that the mode coupling effect is strong enough to
have marked observational consequences \citep{PL00,P01,P03a}.

Polarization evolution of radio waves in pulsar plasma differs
significantly from the evolution in the interstellar medium. In
contrast to the case of Faraday rotation, within the pulsar
magnetosphere the natural waves propagate quasi-transversely with
respect to the magnetic field and have linear polarization. This
rather corresponds to the Cotton-Mouton birefringence (or so
called generalized Faraday rotation, in terms of the paper by
\citealt{KM98}). Another important distinctive feature of
polarization evolution in pulsar magnetosphere is that it takes
place in an essentially inhomogeneous medium. The magnetic field
of a pulsar has approximately dipolar structure, and furthermore,
because of continuity of the plasma flow in the tube of open
magnetic lines, the plasma number density decreases rapidly with
distance. Thus, the character of polarization evolution in pulsar
magnetosphere is quite specific, though the underlying physics of
birefringence is certainly the same.

As a result of the mode coupling effect, the outgoing waves
acquire purely orthogonal elliptical polarization, matching the
empirical representation of superposed elliptical OPMs. It is
important to note that the degree of circular polarization of the
modes and their shift in PA are related to each other, both being
determined by the parameters of the plasma flow in the region of
wave mode coupling \citep{P03a}. Hence, the observed
pulse-to-pulse variations in the ellipticity and PA of the OPMs
can be attributed to the fluctuations in pulsar plasma. Besides
that, the propagation origin of pulsar polarization should imply a
correlation between the values of the ellipticity and PA of the
OPMs at a given pulse longitude. An evidence for such a
correlation has recently been found in \citet{E04} \citep[for more
details see][]{P06}. Thus, the mode coupling effect can account
for a number of important features of the observed single-pulse
polarization.

At the same time, the question as to the origin of
non-orthogonality of polarization modes still remains open. The
observational manifestations of this phenomenon have recently been
reported in a number of papers
\citep[e.g.][]{ES04,Mc04,Rank03,KarIII,KarI}. In the present
paper, we concentrate on a more detailed treatment of polarization
evolution in pulsar magnetosphere, which, in particular, explains
non-orthogonality of the modes.

\subsection{Statement of the problem}
Polarization evolution in pulsar magnetosphere has previously been
considered in the approximation of a superstrong magnetic field.
It means that in the rest frame of the plasma flow the radio
frequency, $\omega^\prime$, is much less than the electron
gyrofrequency, $\omega_H$. In other words, the radius of cyclotron
resonance, where $\omega^\prime =\omega_H$, has been assumed to be
infinitely large. At the conditions relevant to pulsar
magnetosphere, the radius of cyclotron resonance is often somewhat
larger than that of the mode coupling region, but generally these
quantities are of the same order of magnitude \citep[][see also
equation (15) below]{B86}. Therefore it is reasonable to inspect
the role of the cyclotron resonance in the evolution of pulsar
polarization.

In application to pulsars, the cyclotron absorption has been
considered in a number of papers \citep{BS76,LP98,P02,P03b} and
found efficient, especially in case of small pitch-angles of the
absorbing particles. In these papers, it is assumed that the
resonant photons interact with a system of absorbing particles
rather than with the plasma. The main motivation for such an
assumption is that the resonance region lies in the outer
magnetosphere, where the plasma number density is small enough.
Indeed, in case of pulsars, taking into account the plasma effect
on the process of cyclotron absorption introduces only small
corrections to the absorption coefficients of the natural modes
and does not change the total intensities of outgoing radiation
considerably \citep{LP98}. At the same time, cyclotron absorption
in the plasma may markedly affect polarization evolution of radio
waves. The contribution of cyclotron absorption, though not large
quantitatively, may appear comparable to that of the mode coupling
effect, modifying the final polarization of a pulsar drastically.

Given that the regime of geometrical optics is still valid in the
region of cyclotron resonance, the ordinary and extraordinary
waves are absorbed independently, with the absorption coefficients
being slightly different. Note that this difference is purely the
plasma effect, not characteristic of a simple system of absorbing
particles. Beyond the resonance region, the waves propagate in the
weakly magnetized plasma with rapidly decreasing number density
and finally suffer the mode coupling. It should be noted that in
case of weakly magnetized plasma the limiting polarization differs
substantially from that for the superstrong magnetic field.

Given that the natural waves pass through the coupling region
before the cyclotron resonance, in the resonance region each of
the waves presents a coherent mixture of the two natural waves.
Since for these constituents absorption is not identical, the wave
polarization changes considerably. It is important to note that
for the two incident waves polarization evolution is not the same,
since they contain different portions of the ordinary and
extraordinary waves. As a result, polarization states of the
outgoing waves are non-orthogonal.

In the present paper, we concentrate on the analytical
consideration of polarization transfer in pulsar plasma, taking
into account the effect of cyclotron resonance. The plan of the
paper is as follows. In Sect. 2, the main equations are derived,
which describe the evolution of the wave fields in the
inhomogeneous hot plasma embedded in the magnetic field. The basic
numerical estimates are also given there. In Sect. 3, we solve the
equations of polarization transfer in the two limiting cases, when
the resonance region is well below and well above the mode
coupling region, respectively. Section 4 contains the results of
numerical tracing of polarization evolution. In Sect. 5, the
observational consequences of polarization transfer in pulsar
magnetosphere are discussed. Section 6 contains a brief summary. A
statistical model of single-pulse polarization based on the
propagation effects studied will be developed in the forthcoming
paper \citep{P06}.

\section{General theory of polarization evolution in pulsar magnetosphere}
\subsection{Basic equations}
Let radio waves propagate in the ultrarelativistic highly
magnetized  electron-positron plasma of pulsar magnetosphere. The
problem on polarization evolution of the waves in case of
infinitely strong magnetic field has been considered in
\citet{LP99} and \citet{PL00}. The wave propagation in a cold
relativistically streaming plasma embedded in  the magnetic field
of a finite strength has been studied in \citet{P01}. However, in
that paper the plasma number density has been assumed low enough
for the polarization evolution to cease well below the radius of
cyclotron resonance and the wave passage through the resonance has
not been considered. Below we derive the generalized equations
which allow for the cyclotron resonance and are applicable to the
more realistic case of a hot plasma.

The evolution of the wave fields ${\bf E}$ and ${\bf B}$ is
described by the Maxwell's equations: \begin{equation}
\nabla\times {\bf E}=\frac{i\omega}{c}{\bf B}\,,
\end{equation}
\begin{equation}
\nabla\times {\bf B}=-\frac{i\omega}{c}{\bf
E}+\frac{4\pi}{c}\sum_\alpha{\bf j }_\alpha\,.
\end{equation} Here
${\bf j}_\alpha$ is the linearized current density of the
electrons or positrons, summation is over the particle species,
and the time dependence is taken in the form ${\rm e}^{-i\omega
t}$. We are going to trace polarization evolution starting from
the place where refraction is already inefficient and the natural
waves propagate straight. Hence, one can choose a
three-dimensional Cartesian system with the z-axis along the wave
trajectory and the xz-plane being the plane of the field lines of
the dipolar magnetic field of a pulsar at the starting point. Then
all the quantities in equations (1)-(2) depend only on the
z-coordinate.

Since the refractive indices of both natural waves, the ordinary
and extraordinary ones, are close to unity, the wave electric
field can be presented as
\begin{equation}
E_{x,y,z}=\widetilde{E}_{x,y,z}{\rm e}^{i\frac{\omega}{c}z}\,,
\end{equation} where
$\widetilde{E}_{x,y,z}$ vary weakly over the wavelength:
\begin{equation}
\frac{{\rm d}\widetilde{E}_{x,y,z}}{{\rm d}z}\ll
\widetilde{E}_{x,y,z}\frac{\omega}{c}\,.
\end{equation}
All the
other quantities entering equations (1)-(2) can be presented
similarly. Further on the tildes will be omitted. The
representation (3)-(4) corresponds to the case of a weakly
inhomogeneous medium. Typically the scale length for change in the
plasma parameters is about the altitude above the neutron star,
$10^7-10^8$ cm, and the assumption of weak inhomogeneity is
justified. Note that equation (4) requires also the length of the
cyclotron resonance region to be much more than the wavelength.
This is valid for the hot plasma, in which case the wave passes
successively through the resonance with the particles of different
energies.

Substituting equation (1) into equation (2) and making use of
equations (3) and (4) yields: \[\frac{{\rm d}E_x}{{\rm
d}z}+\frac{2\pi}{c}\sum_\alpha j_{\alpha x}=0\,,\]
\begin{equation}
\frac{{\rm d}E_y}{{\rm d}z}+\frac{2\pi}{c}\sum_\alpha j_{\alpha
y}=0\,,
\end{equation} \[E_z+\frac{4\pi i}{\omega}\sum_\alpha
j_{\alpha z}=0\,.\] As can be seen from the above equations,
$E_z\ll E_{x,y}$, that is the waves are practically transverse;
hereafter it is taken that $E_z=0$. Furthermore, for the
components of ${\bf j}_\alpha$ in the first two equations it is
sufficient to use the expressions obtained in the approximation of
a homogeneous medium, where the fields and currents are $\propto
{\rm e}^{i\frac{\omega}{c}z}$, since the corrections for the
medium inhomogeneity are too small (cf. equation (4)). The
conductivity tensor of a hot magnetized homogeneous plasma is well
known \citep[e.g.][]{Mikh75}, however, it is typically written for
the coordinate system with the z-axis along the magnetic field
direction rather than along the wavevector, as is necessary for
our problem. The corresponding transformation of the conductivity
tensor is performed in Appendix. Using those results in equation
(5), we find finally: \[\frac{{\rm d}E_x}{{\rm
d}z}+\frac{i\omega}{2c}\left [Ab_x(E_xb_x+E_yb_y)-BE_x+iGE_y\right
]=0\,,\] \begin{equation} \frac{{\rm d}E_y}{{\rm
d}z}+\frac{i\omega}{2c}\left [Ab_y(E_xb_x+E_yb_y)-BE_y-iGE_x\right
]=0\,.
\end{equation}
Here $b_{x,y,z}$ are the direction cosines of the ambient magnetic
field, \[A\equiv\sum_\alpha\left [\left
[\frac{\omega_{p\alpha}^2f_\alpha(\gamma_\alpha)}
{\gamma_\alpha\omega_\alpha^{\prime
2}}\frac{\omega_H^2}{\omega_H^2-\omega_\alpha^{\prime 2}}\right
]\right ]\,,\] \[B\equiv\sum_\alpha\left [\left
[\frac{\omega_{p\alpha}^2f_\alpha(\gamma_\alpha)\gamma_\alpha\left
(1-\beta_\alpha b_z\right )^2 } {\omega_H^2-\omega_\alpha^{\prime
2}}\right ]\right ]\,,\] \begin{equation} G\equiv\sum_\alpha\left
[\left
[\frac{\omega_{p\alpha}^2f_\alpha(\gamma_\alpha)(q_\alpha/e)(\omega_H/\omega)\left
(\beta_\alpha -b_z\right ) } {\omega_H^2-\omega_\alpha^{\prime
2}}\right ]\right ]\,,
\end{equation}
$\omega_p\equiv\sqrt{\frac{4\pi e^2n}{m_e}}$ is the plasma
frequency, $n$ the number density,
$\omega^\prime\equiv\omega\gamma (1-\beta b_z)$ is the wave
frequency in the frame of the plasma particle moving at a speed of
$\beta\equiv v/c$, $\gamma\equiv\left (1-\beta^2\right )^{-1/2}$
the Lorentz-factor, $q_\alpha=\pm e$, $f(\gamma)$ the distribution
function of electrons or positrons with the normalization $\int
f(\gamma) {\rm d}\gamma\equiv 1$, the double square brackets stand
for the Landau integral: \begin{equation} \left [\left
[\frac{F(\gamma)}{\omega_H^2-\omega^{\prime 2 }}\right ]\right
]={\rm v.p.}\int\frac{F(\gamma){\rm d
}\gamma}{\omega_H^2-\omega^{\prime 2 }}+\pi
i\int\frac{\delta(\omega_H-\omega^\prime)F(\gamma){\rm d
}\gamma}{\omega_H+\omega^\prime} \,,
\end{equation} the first of
the above integrals is taken in the principal value sense.

Equations (6)-(8) describe polarization evolution of the waves in
the hot magnetized weakly inhomogeneous plasma allowing for the
effect of cyclotron resonance. These equations coincide with
equation (12) in \citet{P01} in case of cold plasma, with the
distribution function $f(\gamma)=\delta(\gamma-\gamma_0)$, if only
a part of the wave trajectory well below the resonance radius is
considered.

It should be noted that equation (6) incorporates the current
density in the limit of small pitch-angles, $\psi$, of the plasma
particles (for more detail see Appendix):
\begin{equation}
\frac{\omega^\prime}{\omega_H}\frac{\psi}{\theta}\ll
1\,,
\end{equation}
where $\theta\approx\sqrt{b_x^2+b_y^2}$ is the
angle between the wavevector and the ambient magnetic field. In
the resonance region, equation (9) means that $\psi\ll\theta$.
Since in pulsar magnetosphere the optical depth to cyclotron
absorption is typically large (cf. equation (12) below), the
absorbing particles can increase their pitch-angles significantly
and the approximation $\psi\ll\theta$ can be broken
\citep{LP98,P02,P03b}. This especially concerns low enough radio
frequencies which meet the resonance condition at higher altitudes
above the neutron star and thus interact with the particles whose
distribution function has already evolved because of absorption of
the higher-frequency radio photons. At the same time, absorption
by the particles with $\psi\ll\theta$ is still characteristic of
pulsar magnetosphere, and in the present paper we concentrate on
this case of small pitch-angles.

\subsection{Numerical estimates}
As can be seen from the wave equation (6), the terms containing
$B$ can be excluded by introducing the substitution
$\widehat{E}_{x,y}=E_{x,y}{\rm e}^{i\frac{\omega}{2c}\int B{\rm d
}z}$, i.e. these terms act to decrease both field components
identically. With the common expression for the total intensity,
$I=E_xE_x^*+E_yE_y^*$ (where the asterisk stands for complex
conjugation), one can easily recognize the coefficient of
cyclotron absorption: \begin{equation} \mu =\frac{\omega}{c}\Im
B\,.
\end{equation}
Using equations (7) and (8), it can be reduced to the form
\begin{equation}
\mu =\frac{2\pi^2ne^2}{mc\omega}f\left
(\frac{\omega_H}{\omega\theta^2/2}\right )\,,
\end{equation} which coincides with that known for the coefficient
of absorption by the system of particles in vacuum (e.g. equation
(4.13) in \citealt{LP98}). Above it is taken into account that the
wave propagation is quasi-transverse with respect to the ambient
magnetic field, $\theta\ll 1/\gamma$. For the conditions relevant
to pulsar magnetosphere, the optical depth to cyclotron
absorption, $\Gamma_c=\int\mu {\rm d}z$, is estimated as follows
(e.g. equation (2.8) in \citealt{LP98}):
\begin{equation}
\Gamma_c=\frac{0.4\kappa_2B_{\star
12}^{3/5}\sin^{4/5}\vartheta}{\left (P^3\nu_9\gamma_{1.5}\right
)^{3/5} }\,.
\end{equation}
Here $\kappa$ is the plasma
multiplicity factor, $\kappa_2\equiv\frac{\kappa}{10^2}$,
$B_\star$ the magnetic field strength at the surface of the
neutron star, $B_{\star 12}\equiv\frac{B_\star}{10^{12}\,{\rm
G}}$, $\vartheta$ the angle between the rotational and magnetic
axes of a pulsar, $P$ the pulsar period, $\nu$ the radio
frequency, $\nu_9\equiv\frac{\nu}{10^9\,{\rm Hz}}$,
$\gamma_{1.5}\equiv\frac{\gamma}{10^{1.5}}$. One can see that at
low enough frequencies the absorption depth can be significant,
especially for the short-period pulsars, $P\sim 0.1$ s. Below we
shall no longer be interested in the consequent intensity decrease
(equal for both natural waves) and concentrate on examining the
normalized Stokes parameters with an eye to tracing polarization
evolution of the waves.

In the wave equation (6), the terms containing $G$ allow for the
plasma gyrotropy: if the distribution functions of electrons and
positrons are identical, $G\equiv 0$. The problem on the plasma
motion in the electromagnetic fields of pulsar magnetosphere is
very complex and the self-consistent solution has not been found
yet. Thus, the question on the net current density in the
magnetosphere is still open. The simplest possibility is that the
electrons and positrons differ only in the number densities, with
the difference being of order of the Goldreich-Julian number
density, i.e. $\Delta n/n\sim 1/\kappa$. Then the relative
contribution of the plasma gyrotropy to polarization evolution can
be estimated as $\Re G/\left [\left (b_x^2+b_y^2\right )\Re
A\right ]\sim\kappa^{-1}(\theta\gamma)^2\omega^\prime/\omega_H\sim
0.1\kappa_2^{-1}\theta_{-1}^2\gamma_{1.5}^2\omega^\prime/\omega_H$.
In the resonance region, it can become significant, at least for
the rays of a specific geometry. However, keeping in mind that
true form of $G$ is unknown, we leave out the plasma gyrotropy and
the resultant ellipticity of the natural waves. Our aim is to
study polarization evolution of the linearly polarized natural
waves and, in particular, to demonstrate how the ellipticity
arises and changes purely on account of wave mode coupling and
cyclotron absorption. An opposite approach has recently been
developed by \citet{Mel04b} and \citet{Mel04a}, who investigated
the characteristics of the elliptically polarized natural waves
ignoring the above mentioned effects of polarization evolution.
The propagation of the elliptical natural waves through the region
of cyclotron resonance has been considered in \citet{MelrPhRev04}.

In the present paper, we neglect the terms  $B$ and $G$ in
equation (6) and concentrate solely on $A$. Note that $\Re A$
describes wave mode coupling, while $\Im A$ corresponds to
cyclotron absorption. In the resonance region, these two
contributions are roughly of the same order.

Taking into account that the magnetic field strength decreases
with distance from the neutron star as $z^{-3}$, one can estimate
the radius of cyclotron resonance, where
$\omega\gamma\theta^2/2=\omega_H$, for the particles with some
characteristic Lorentz-factor $\gamma$ as
\begin{equation}
\frac{z_c}{r_L}=0.55P^{-1}B_{\star
12}^{1/3}\nu_9^{-1/3}\gamma_{1.5}^{-1/3}\theta_{-1}^{-2/3}\,.
\end{equation} Here $r_L\equiv 5\cdot 10^9P$ cm is the light
cylinder radius, $\theta_{-1}\equiv\theta/0.1$. The altitude of
the mode coupling region, $z_p$ is another basic quantity of the
problem considered. Proceeding from the definition of $z_p$,
\[\frac{2\omega_p^2}{\gamma^3\omega^2\theta^2}\frac{\omega}{c}z_p=1\]
\citep[for more detail see, e.g. ][]{P03a}, one can obtain the
following estimate:
\begin{equation}\frac{z_p}{r_L}=0.18P^{-3/2}\gamma_{1.5}^{-3/2}\kappa_2^{1/2}B_{\star
12}^{1/2}\nu_9^{-1/2}\theta_{-1}^{-1}\,.
\end{equation} Hence, the
altitudes of the regions of pronounced polarization evolution are
in the ratio
\begin{equation}
\frac{z_p}{z_c}=0.33P^{-1/2}\gamma_{1.5}^{-7/6}\theta_{-1}^{-1/3}B_{\star
12}^{1/6}\nu_9^{-1/6}\kappa_2^{-1/2}\,.
\end{equation}
One can see
that for the parameters relevant to pulsars this ratio can take
the values more and less than unity. These two possibilities imply
qualitatively different scenarios of polarization evolution. In
the next Section, we dwell on the analytical treatment of the wave
equation in the limiting cases $z_p/z_c\gg 1$ and $z_p/z_c\ll 1$,
and find out general features of polarization behaviour in these
two scenarios.

\section{Analytical treatment of polarization evolution}
In terms of the Stokes parameters, \[I=E_xE_x^*+E_yE_y^*\,,\]
\[q=E_xE_x^*-E_yE_y^*\,,\] \[u=E_xE_y^*+E_yE_x^*\,,\]
\[v=i(E_xE_y^*-E_yE_x^*)\,,\] where the asterisk stands for complex
conjugation, the wave equations (6) are written as \[\frac{{\rm
d}I}{{\rm
d}z}=2A_2b_xb_yu+A_2(b_x^2-b_y^2)q+A_2(b_x^2+b_y^2)I\,,\]
\[\frac{{\rm d}q}{{\rm
d}z}=2A_1b_xb_yv+A_2(b_x^2-b_y^2)I+A_2(b_x^2+b_y^2)q\,,\]
\[\frac{{\rm d}u}{{\rm
d}z}=2A_2b_xb_yI-A_1(b_x^2-b_y^2)q+A_2(b_x^2+b_y^2)u\,,\]
\begin{equation}
\frac{{\rm d}v}{{\rm
d}z}=-2A_1b_xb_yq+A_1(b_x^2-b_y^2)u+A_2(b_x^2+b_y^2)v\,.
\end{equation}
Here $A_1\equiv\frac{\omega}{2c}\Re A$,
$A_2\equiv\frac{\omega}{2c}\Im A$ and, in accordance with the
above discussion, the terms containing $B$ and $G$ are left aside.
The last terms in equations (16) can be excluded using the
substitution $\widetilde{s}_m=s_m\exp[-\int A_2(b_x^2+b_y^2){\rm
d}z]$, $m=1,...,4$, where $s_m$ stands for the $m$-th Stokes
parameter. Our primary interest is the evolution of the normalized
Stokes parameters, so hereafter we consider the quantities
$\widetilde{s}_m$ and omit the tildes.

Introducing the normalized coordinate $w\equiv z_p/z$, one can
present equations (16) in the following form: \[\frac{{\rm
d}I}{{\rm d }w}=-wF_2q-2\mu F_2u\,,\] \[\frac{{\rm d}q}{{\rm d
}w}=-wF_2I-2\mu F_1v\,,\] \[\frac{{\rm d}u}{{\rm d }w}=wF_1v-2\mu
F_2I\,,\]
\begin{equation}
\frac{{\rm d}v}{{\rm d }w}=-wF_1u+2\mu F_1q\,,
\end{equation}
where \[F_1\equiv {\rm v.p.}\int \frac{f(\gamma){\rm
d}\gamma}{(\gamma /\gamma_0)^3[1-(\gamma
/\gamma_0)^2(z/z_c)^6]}\,,\] \[F_2\equiv\frac{\pi}{2}\left
(z/z_c\right )^6\gamma_0f\left [\left (z/z_c\right
)^{-3}\gamma_0\right ]\,,\] $\gamma_0$ is the characteristic
Lorentz-factor of the plasma particles with the distribution
function $f(\gamma)$ and $z_c$ is the characteristic altitude of
the resonance region defined by the condition
$\omega_H(z_c)=\omega\gamma_0\theta^2/2$. Due to continuity of the
plasma flow in the tube of open magnetic lines, the plasma number
density $n\propto B\propto z^{-3}$. The radio waves are emitted in
the plane of magnetic lines and hence initially $b_y=0$. As the
waves propagate in the rotating magnetosphere, they leave the
plane of magnetic lines, so that $b_y$ slightly increases along
the trajectory, $b_y\sim z/r_L$. For the sake of simplicity, in
the present paper this angle is assumed to remain much less than
the wavevector tilt in the plane of magnetic lines, $b_x$:
$\mu\equiv(b_y/b_x)_{z_p}\ll 1$.

Strictly speaking, the wave equations (6) correspond to the frame
corotating with the neutron star; for the observer's frame the
corrections for rotational aberration should be included
\citep[cf., e.g. ][]{LP99}. Similarly to $b_y$, these corrections
are also $\sim z/r_L$ and depend on the angle between the
rotational and magnetic axes of the pulsar, the observational
geometry and pulse longitude. However, here we are interested only
in the principal features of polarization evolution, so we do not
specify the geometrical parameters and assume that $\mu$ already
includes the factor of order unity which allows for rotational
aberration.

It is convenient to assume that the particle Lorentz-factors lie
within a finite interval $[\gamma_1 ,\gamma_2]$. Then the
resonance region is also finite, with the boundaries given by the
conditions $\omega_H(z_{c_{1,2}})=\omega\gamma_{1,2}\theta^2/2$,
and the wave trajectory is divided into the three segments lying
below, inside and above the resonance region, respectively. Let us
consider the polarization transfer in each of these regions
separately.

\subsection{The region before cyclotron resonance}
Below the region of cyclotron resonance, $F_2\equiv 0$, and the
set of equations (17) is markedly simplified. For the natural
waves of pulsar plasma, the initial conditions read: $q_0/I_0=\pm
1$, $u_0=v_0=0$, where the upper and lower signs correspond to the
ordinary and extraordinary waves, respectively. It is easy to see
that if $\mu \ll 1$, in the region considered the quantities
$\vert u\vert$ and $\vert v\vert$ remain $\sim \mu$, while $\vert
q\vert\approx 1$. Thus, equation (17) can be reduced to the form
\[\frac{{\rm d}u}{{\rm d}w}=wF_1v\,,\]
\begin{equation}
\frac{{\rm d}v}{{\rm d}w}=-wF_1u+2\mu F_1\,.
\end{equation}
Substituting the second of the above equations into the first one,
one can find the solution:
\begin{equation}
u_{\rm I}=\pm 2\mu\Im
R_{\rm I}\,,\quad v_{\rm I}=\mp 2\mu\Re R_{\rm I}\,,
\end{equation}
where
\begin{equation}
R_{\rm I}\equiv {\rm e}^{-i\int^{w}F_1w^\prime{\rm
d}w^\prime}\int_{w}^{\infty}{\rm e}^{i\int^{w^\prime}F_1w{\rm
d}w}F_1{\rm d}w^\prime\,.
\end{equation}
The above solution
differs substantially for the cases $w_1\gg 1$ and $w_1\ll 1$,
where $w_1$ is the lower boundary of the resonance region. It is
convenient to introduce the variable $x\equiv w/\eta=z_c/z$, which
remains of order unity, and estimate the integral in equation (20)
at $\eta\gg 1$ and $\eta\ll 1$.

Given that $\eta\gg 1$, we have
\begin{equation}
u_{\rm I}\approx\pm\frac{2\mu}{\eta x}\,,\quad v_{\rm
I}\approx\mp\frac{2\mu}{F_1\eta^3x^3}\,,
\end{equation}
which is a
common solution in the regime of geometrical optics. In the
opposite limit, $\eta\ll 1$, the waves pass through the region of
wave mode coupling. It is reasonable to present the integral (20)
as a sum of the two integrals over the intervals $\infty
>w>\sqrt{\eta}$ and $\sqrt{\eta}>w>\eta u_1$. The boundary between
the intervals, $u_a=1/\sqrt{\eta}\,,w_a=\sqrt{\eta}$, is chosen so
that $w$ is small enough for the wave mode coupling to be
efficient only within the first interval and at the same time it
is far enough from the resonance region. In the first interval,
the solution has the form \[ u_{{\rm
I}_a}\approx\pm\mu\sqrt{\pi}(1-w^2/2)\,,\]
\begin{equation}
v_{{\rm
I}_a}\approx\mp\mu\sqrt{\pi}(1+w^2/2)\pm2\mu w\,.
\end{equation}
To the first order in $\eta$, for the second interval we have the
following solution: \[ u_{\rm I}=\pm\mu\sqrt{\pi}\,,\]
\begin{equation}v_{\rm I }=\mp\mu\sqrt{\pi}\pm
2\mu\eta\int_{1/\sqrt{\eta}}^{u_1}F_1{\rm d }u\pm
2\mu\sqrt{\eta}\,.
\end{equation}
Comparing the solutions (21) and
(23) obtained for the limiting cases $\eta\gg 1$ and $\eta\ll 1$,
respectively, one can see that the waves coming to the resonance
region can have substantially distinct polarization
characteristics.

\subsection{The resonance region}
In the resonance region, the set of equations (17) has the form
\[\frac{{\rm d}I}{{\rm d}x}=-\eta^2xF_2q\,,\] \[\frac{{\rm d}q}{{\rm
d}x}=-\eta^2xF_2I\,,\] \[\frac{{\rm d}u}{{\rm
d}x}=\eta^2xF_1v-2\mu\eta F_2I\,,\]
\begin{equation}
\frac{{\rm d}v}{{\rm d}x}=-\eta^2xF_1u+2\mu\eta
F_1q\,,
\end{equation}
where the terms $\sim \mu^2$ are omitted. For the first two
equations, one can find the following solutions: \[q_{\rm
o}=I_{\rm o}=\exp\left [\eta^2\int_x^{x_1}F_2x^\prime{\rm
d}x^\prime\right ]\,,\] \begin{equation}q_{\rm e}=-I_{\rm
e}=-\exp\left [-\eta^2\int_x^{x_1}F_2x^\prime{\rm d}x^\prime\right
]\,,\end{equation} where the subscripts "o" and "e" correspond to
the original ordinary and extraordinary waves, respectively. One
can see that the two types of waves evolve in a different manner.
It should be kept in mind, however, that actually the ordinary
mode intensity does not increase, since it contains an additional
factor $\exp[-\eta^2\int_x^{x_1}F_2x^\prime{\rm d}x^\prime]$
omitted after equation (16). Thus, the ordinary wave is absorbed
by the plasma just as by the system of particles in vacuum (the
absorption coefficient is given by equation (11)), whereas the
extraordinary wave is absorbed somewhat more efficiently.

Using equations (25) in the last two equations of the set (24),
one can obtain: \[v_{\rm II}=\pm\frac{2\mu}{\eta}\Im R_{\rm
II}+v_0\cos\left (\eta^2\int_{x_1}^xF_1x^\prime{\rm
d}x^\prime\right )-\left [u_0\mp\frac{2\mu}{\eta x_1}\right
]\sin\left (\eta^2\int_{x_1}^xF_1x^\prime{\rm d}x^\prime\right
)\,,\] \begin{equation} u_{\rm II}=\pm\frac{2\mu}{\eta}\Re R_{\rm
II}\pm\frac{2\mu}{\eta x}\exp\left
[\pm\eta^2\int_x^{x_1}F_2x^\prime{\rm d}x^\prime\right ]
+v_0\sin\left (\eta^2\int_{x_1}^xF_1x^\prime{\rm d}x^\prime\right
)+\left [u_0\mp\frac{2\mu}{\eta x_1}\right ]\cos\left
(\eta^2\int_{x_1}^xF_1x^\prime{\rm d}x^\prime\right )\,,
\end{equation}
where \[R_{\rm II}\equiv\exp\left [-i\eta^2\int^xF_1x^\prime{\rm
d}x^\prime\right ]\times \int_{x_1}^x\frac{\exp\left
[i\eta^2\int^{x^\prime}F_1x{\rm d }x\right ]\exp\left
[\mp\eta^2\int_{x_1}^{x^\prime} F_2x{\rm d}x\right
]}{x^{\prime^2}}{\rm d}x^\prime\,,\] \[v_0\equiv v_{\rm
I}(x_1)\,,\quad u_0\equiv u_{\rm I}(x_1)\,.\]

In the limit $\eta\gg 1$, we are interested only in the evolution
of the ordinary wave, since the extraordinary one is severely
absorbed. The asymptotic treatment of equation (26) yields:
\begin{equation}
(u/I)_{\rm II}=\frac{2\mu}{\eta x}\,,\quad (v/I)_{\rm
II}=-\frac{2\mu}{F_1\eta^3x^3}\,.
\end{equation} This solution
still coincides with that known for the regime of geometrical
optics in the infinitely strong magnetic field (cf. equation
(21)). Thus, if in the resonance region the wave propagation obeys
geometrical optics, i.e. if $\eta\gg 1$, the cyclotron absorption
does not affect polarization state of the wave and acts only to
suppress the total intensity.

In the limit $\eta\ll 1$, equation (26) is reduced to the
following form: \[u_{\rm
II}=\pm\mu\sqrt{\pi}-2\mu\eta\int_{x_1}^xF_2{\rm d }x^\prime\,,\]
\begin{equation}
v_{\rm
II}=\mp\mu\sqrt{\pi}\pm2\mu\eta\int_{1/\sqrt{\eta}}^xF_1{\rm d
}x^\prime\pm 2\mu\sqrt{\eta}\,.
\end{equation}
In contrast to the
case of geometrical optics, in case $\eta\ll 1$ the cyclotron
absorption changes the polarization state of the waves (cf.
$u_{\rm I}$ and $u_{\rm II}$). This can be understood as follows.
The waves are subject to wave mode coupling well before the
resonance region and become a coherent mixture of the two natural
waves. Since these constituents are absorbed not identically, the
resultant polarization is modified. For the original ordinary and
extraordinary waves, the amplitude ratio of the entering natural
waves is essentially distinct, so they suffer different
polarization evolution (see equation (28)).

\subsection{The region beyond cyclotron resonance}
As soon as the waves go out of the resonance region, their
intensities stop changing, whereas the polarization still evolves.
The final polarization state of the waves at infinity is as
follows: \[ u_f=2\mu\eta\Im R_{\rm III}+u_0\cos\left
[\eta^2\int_0^{x_2}F_1x{\rm d}x\right ]-v_0\sin\left
[\eta^2\int_0^{x_2}F_1x{\rm d}x\right ]\,,\]
\begin{equation} v_f=-2\mu\eta\Re R_{\rm III}+u_0\sin\left
[\eta^2\int_0^{x_2}F_1x{\rm d}x\right ]+v_0\cos\left
[\eta^2\int_0^{x_2}F_1x{\rm d}x\right ]\,,
\end{equation}
where \[R_{\rm III}\equiv{\rm e}^{-i\eta^2\int^0F_1x{\rm d
}x}\int_0^{x_2}{\rm e}^{i\eta^2\int^xF_1x^\prime{\rm
d}x^\prime}F_1{\rm d}x\,,\] \[u_0\equiv u_{\rm II}(x_2)\,,\quad
v_0\equiv v_{\rm II}(x_2)\,.\]

In case $\eta\ll 1$, equation (29) is reduced to
\[u_f=u_0=\pm\mu\sqrt{\pi}-2\mu\eta\int_{x_1}^{x_2}F_2{\rm
d}x\,,\]
\begin{equation}
v_f=\mp\mu\sqrt{\pi}\pm 2\mu\eta\int_{1/\sqrt{\eta}}^0F_1{\rm
d}x\pm2\mu\sqrt{\eta}\,.
\end{equation}
Hence, the wave
polarization does not suffer marked changes beyond the resonance
region.

Given that $\eta\gg 1$, in the region under consideration, the
wave polarization evolves drastically because of wave mode
coupling. In equation (29), the addends containing $u_0$ and $v_0$
compensate for the first terms of the expansion of $R_{\rm III}$
in power of $\eta^{-2}$, and the main contribution comes from the
stationary point of $R_{\rm III}$. The final polarization takes
the form
\[(u/I)_f=2^{5/8}\mu\eta^{-3/4}\Gamma(7/8)\cos(\pi/16)\,,\]
\begin{equation}
(v/I)_f=2^{5/8}\mu\eta^{-3/4}\Gamma(7/8)\sin(\pi/16)\,,
\end{equation}
where $\Gamma(\cdot)$ is the gamma-function and it is taken into
account that far from the resonance region $F_1\approx -u^6$,
independently of the detailed form of the distribution function
$f(\gamma)$.

Equations (30) and (31) describe the final polarization of the
waves in cases when the region of wave mode coupling lies much
lower and much higher than the region of cyclotron resonance,
respectively. It should be noted that in the weak magnetic field
polarization evolution as a result of wave mode coupling differs
essentially from that in the superstrong magnetic field. First of
all, the resultant circular polarization has opposite signs (cf.
equations (30) and (31)). Besides that, in the weak field, the
evolution is less pronounced (note the factors $\eta^{-3/4}$ in
equation (31)).

\section{Numerical simulation}
Now let us turn to numerical simulation of polarization transfer
in pulsar plasma based on equation (16). First of all, we
introduce a simplified distribution function of the plasma
particles in the form of a triangle:
\begin{equation}
f(\gamma)=\left\{\begin{array}{lr}\frac{\displaystyle\gamma
-\gamma_0(1-a) }{\displaystyle
a^2\gamma_0^2}\,,&\quad\gamma_0(1-a)\leq\gamma\leq\gamma_0\,,\\
\frac{\displaystyle\gamma_0(1+a)-\gamma}{\displaystyle
a^2\gamma_0^2}\,,&\quad\gamma_0\leq \gamma\leq\gamma_0(1+a)\,.
\end{array}\right.
\end{equation}
Figure 1 shows the functions
$F_1(z_c/z)$ and $F_2(z_c/z)$, which incorporate equation (32),
for the two values of the width of $f(\gamma)$: $a=0.1$ and
$a=0.5$.

According to the approximate solutions (30) and (31), polarization
evolution should be most pronounced at $\eta\approx 1$. In Fig. 2,
we present the numerically simulated evolution of the normalized
Stokes parameters at $\eta =1$. One can see that the changes are
indeed marked and are not identical for the two types of waves.

The final values of the normalized Stokes parameters $u/I$ and
$v/I$ as functions of $\eta$ at different $\mu$ are plotted in
Figs. 3a and b, respectively. The analytical approximations given
by equations (30) and (31) are shown in asterisks (where
appropriate). The polarization parameters orthogonal to those of
the original ordinary mode are shown by dotted lines for
comparison. As can be seen from Fig. 3, at $\eta\sim 1$ the modes
are markedly non-orthogonal. Note also the switch of the sign of
$v$ at $\eta\approx 1$.

At large enough $\eta$, the extraordinary mode abruptly turns into
the ordinary one. This can be understood as follows. Generally
speaking, the natural modes are completely independent only
infinitely far from the region of wave mode coupling, whereas for
finite $\eta$ in the resonance region each of the waves contains
already a small admixture of the other mode. For large enough
$\eta$, the extraordinary component is absorbed much more
efficiently than the ordinary one, so that even in the original
extraordinary wave only the ordinary component survives. Note,
however, that this regime can hardly be observed, because the
ordinary wave intensity is also strongly suppressed by cyclotron
absorption.

\section{Discussion}
As is found above, polarization evolution of radio waves in pulsar
magnetosphere can be substantial. The wave polarization varies
because of wave mode coupling and cyclotron absorption, the change
being most pronounced if the regions of the coupling and resonance
approximately coincide.

In cases when the resonance region lies well below and well above
the region of wave mode coupling ($\eta\gg 1$ and $\eta\ll 1$,
respectively), the resultant polarization is essentially distinct.
First of all, this distinction is caused by the different
character of the wave mode coupling in the approximations of the
weak and strong magnetic field. In the former case, polarization
evolution is weaker and the sign of circular polarization is
different.

For $\eta\gg 1$ and $\eta\ll 1$, the consequences of cyclotron
absorption are also distinct. Given that $\eta\gg 1$, the waves
passing through the resonance region still obey the geometrical
optics approximation and propagate independently. Then the
cyclotron absorption affects only their intensities. In contrast
to the cyclotron absorption by a system of particles, in the
plasma the wave intensities are suppressed not identically: the
extraordinary wave is absorbed more efficiently, the difference
strongly increasing with $\eta$. The numerical calculation shows
that even at $\eta=1$ ($\mu=0.3$, $a=0.1$) the outgoing
intensities differ by a factor of about 4.

In case $\eta\ll 1$, the cyclotron absorption acts mainly to
change the wave polarization. The waves entering the resonance
region present already a coherent mixture of the two types of
natural waves, and since these constituents are absorbed not
identically, the resultant polarization changes. For the original
ordinary and extraordinary waves, polarization evolution on
account of cyclotron absorption is distinct, so that they become
non-orthogonal.

The features of polarization transfer listed above are believed to
have a number of observational consequences. For example, the
pulse-to-pulse fluctuations of $\eta$ around $\eta\approx 1$ can
cause the sense reversals of the circular polarization at a given
pulse longitude \citep[see also][]{MelrPhRev04}, which are really
observed \citep[e.g. ][]{KarI}. It should be kept in mind,
however, that this requires too large fluctuations in the plasma
parameters; besides that, on both sides of $\eta =1$, the absolute
values of $v$ are markedly different, while very often the
observed sense reversals do not change the absolute value of the
circular polarization essentially.

Generally speaking, pulsar radiation presents a mixture of the two
types of waves with nearly orthogonal polarizations and comparable
randomly varying intensities, so that it is partially depolarized.
The cyclotron absorption, especially at large enough $\eta$,
introduces a systematic difference in the randomly varying mode
intensities. Then the ordinary wave should prevail on average, in
which case the degree of polarization should be markedly enhanced.
Large $\eta$ do not seem typical of the pulsars observed, however,
some pulsars with exceptionally high polarization indeed exist.
The Vela presents a well-known example of such a pulsar. As a
rule, in these cases the same polarization mode dominates
throughout the average pulse and in most of the single pulses. It
is interesting to note that a few strongly polarized pulsars have
the total-intensity profiles classified as partial cones
\citep[e.g. ][]{LM88}, which is an additional hint at a
significant cyclotron absorption in these objects.

Unfortunately, the actual value of the position angle of linear
polarization is not determined in observations and remains
unknown, so that direct observations tell nothing as to whether
the ordinary or extraordinary mode dominates in the strongly
polarized pulsars. Even in case of the Vela, where the X-ray
observations of the wind nebula give some new insights into the
pulsar geometry, this question has no firm answer. As for the
theoretical predictions, cyclotron absorption strongly favours the
dominance of the ordinary waves. Let us note in passing that in
case of resonant absorption by the particles with large
pitch-angles, which is a very probable regime for the Vela, the
ordinary waves should also prevail \citep{P02,P03b}.

As can be seen from equation (15), $\eta$ is a very weak function
of radio frequency. Obviously, just for this reason the majority
of pulsars are detectable over a wide frequency range. Let us
speculate, however, that if at the intermediate frequencies
$\eta\sim 1$, at low frequencies $\eta$ is larger and the degree
of polarization can substantially increase, which is in line with
the observational results below 100 MHz \citep{Sul00}. At high
frequencies $\eta <1$ and the polarization modes can be markedly
non-orthogonal, in which case the degree of circular polarization
can be enhanced. The increase of circular polarization at high
frequencies is indeed found in the average profiles of a number of
pulsars \citep[e.g.][]{KarV}.

A possible non-orthogonality of polarization modes on account of
cyclotron absorption has a number of important implications for
the statistical model of the individual-pulse polarization of
pulsars. (This issue will be considered in detail in the
forthcoming paper \citealt{P06}). Given that the observed
superposed orthogonal polarization modes are associated with the
natural waves of pulsar plasma, they should be orthogonal by
definition. The process of wave mode coupling also preserves the
orthogonality of the waves. To the best of our knowledge,
cyclotron absorption is the only way for the {\it superposed}
modes to become non-orthogonal. Note, however, that the
non-orthogonal states of the observed {\it sum} of the modes in
different single pulses can be attributed to the time-dependent
action of propagation effects.

\section{Conclusions}
We have studied polarization transfer in the hot magnetized plasma
of pulsars. In the present consideration, we have assumed the
non-gyrotropic plasma, with the identical distributions of the
electrons and positrons, and the small pitch-angles of the
particles. Proceeding from the Maxwell's equations, we have
derived the set of equations describing the evolution of the
Stokes parameters of the original linearly polarized natural
waves. These equations have been solved analytically and the
results have been confirmed by numerical calculations.

The polarization evolution of the waves has been found
significant. The polarization characteristics change on account of
the wave mode coupling and cyclotron absorption. In cases when the
region of coupling lies well above and well below the resonance
region, the resultant polarization is qualitatively distinct. If
the waves pass through the region of wave mode coupling first,
they acquire the elliptical polarizations purely orthogonal at the
Poincare sphere. Further on, in the resonance region, they become
non-orthogonal. If the waves enter the resonance region before
coupling, cyclotron absorption does not affect their polarization
states and suppresses the total intensities only, the
extraordinary wave being absorbed somewhat more efficiently.
Further on the waves suffer mode coupling in the limit of weak
magnetic field, which differs from that in the strong field.
Firstly, the total change of polarization parameters is much less.
Besides that, the resultant circular polarization has the opposite
sense.

The observational consequences of polarization transfer in pulsar
magnetosphere can be summarized as follows. Because of cyclotron
absorption, at large enough $\eta$ one mode can markedly dominate
another one, in which case one can expect strongly polarized
profiles, with the same mode dominating throughout the average
pulse and in most of the individual pulses. This is indeed
characteristic of several pulsars. In some cases, the
total-intensity profiles of the strongly polarized pulsars are
classified as partial cones, with one of the conal components
being absent. This seems to be an additional argument in favour of
strong cyclotron absorption in these pulsars. At present it is not
known exactly what of the natural waves actually dominates in the
strongly polarized pulsars. Our consideration favours the
dominance of the ordinary mode.

If one consider $\eta$ as a function of radio frequency, with
$\eta\approx 1$ at the intermediate frequencies, in the
low-frequency range $\eta\geq 1$ and one can expect strong
polarization of pulsar profiles, whereas at high enough
frequencies there may be an increase of the degree of circular
polarization because of non-orthogonality of the modes. The
non-orthogonality of the outgoing waves is of a crucial importance
for the statistical model of the individual-pulse polarization.
This will be studied in detail in the forthcoming paper.

\section*{Acknowledgements}
This research is in part supported by INTAS Grant No.~03-5727 and
the Grant of the President of Ukraine (the project No.~GP/F8/0050
of the State Fund for Fundamental Research of Ukraine).

\clearpage \setcounter{figure}{0}
\begin{figure*}
\input epsf
\def\epsfsize#1#2{0.6#1} \centerline{\epsfbox{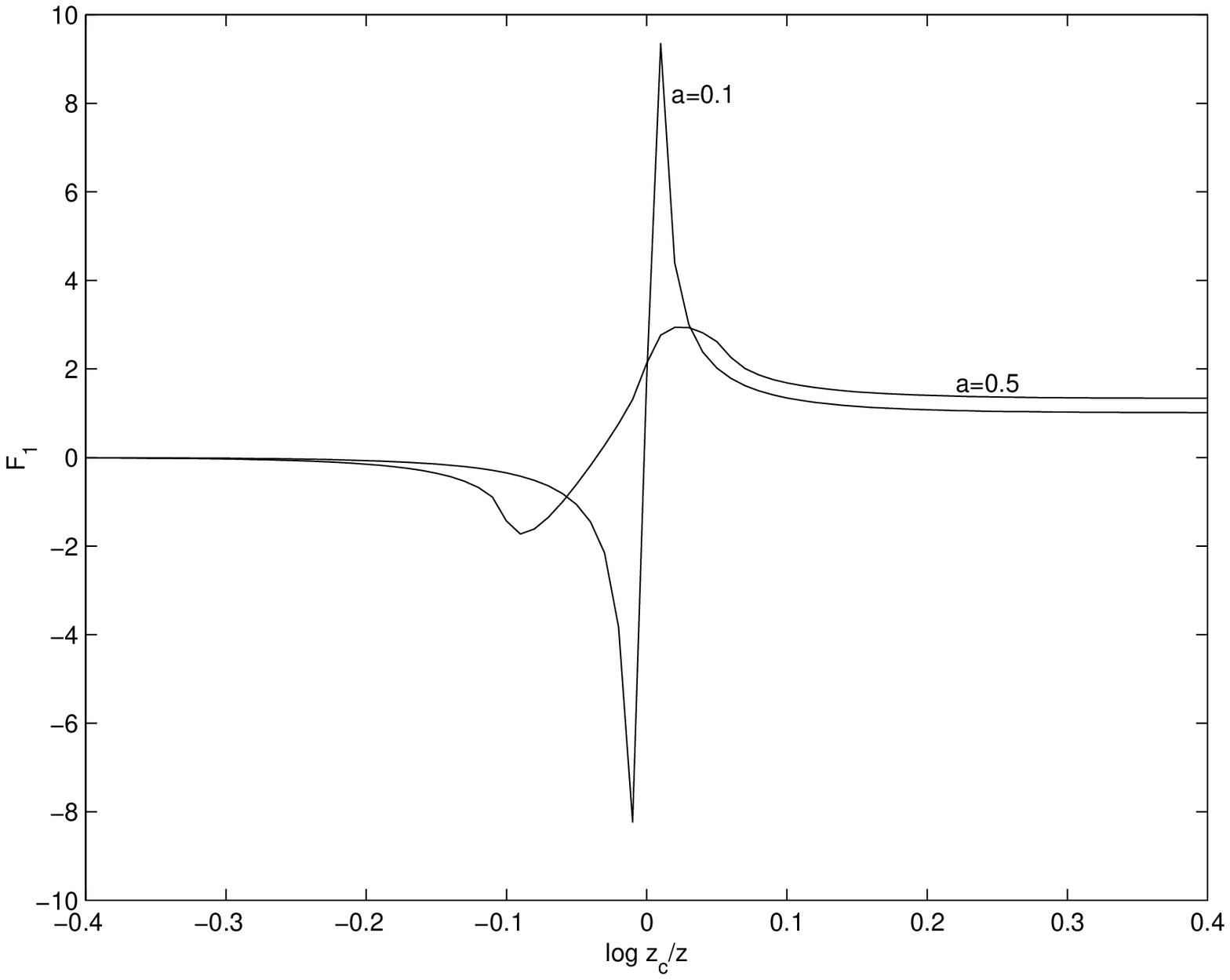}}
\input epsf
\def\epsfsize#1#2{0.6#1} \centerline{\epsfbox{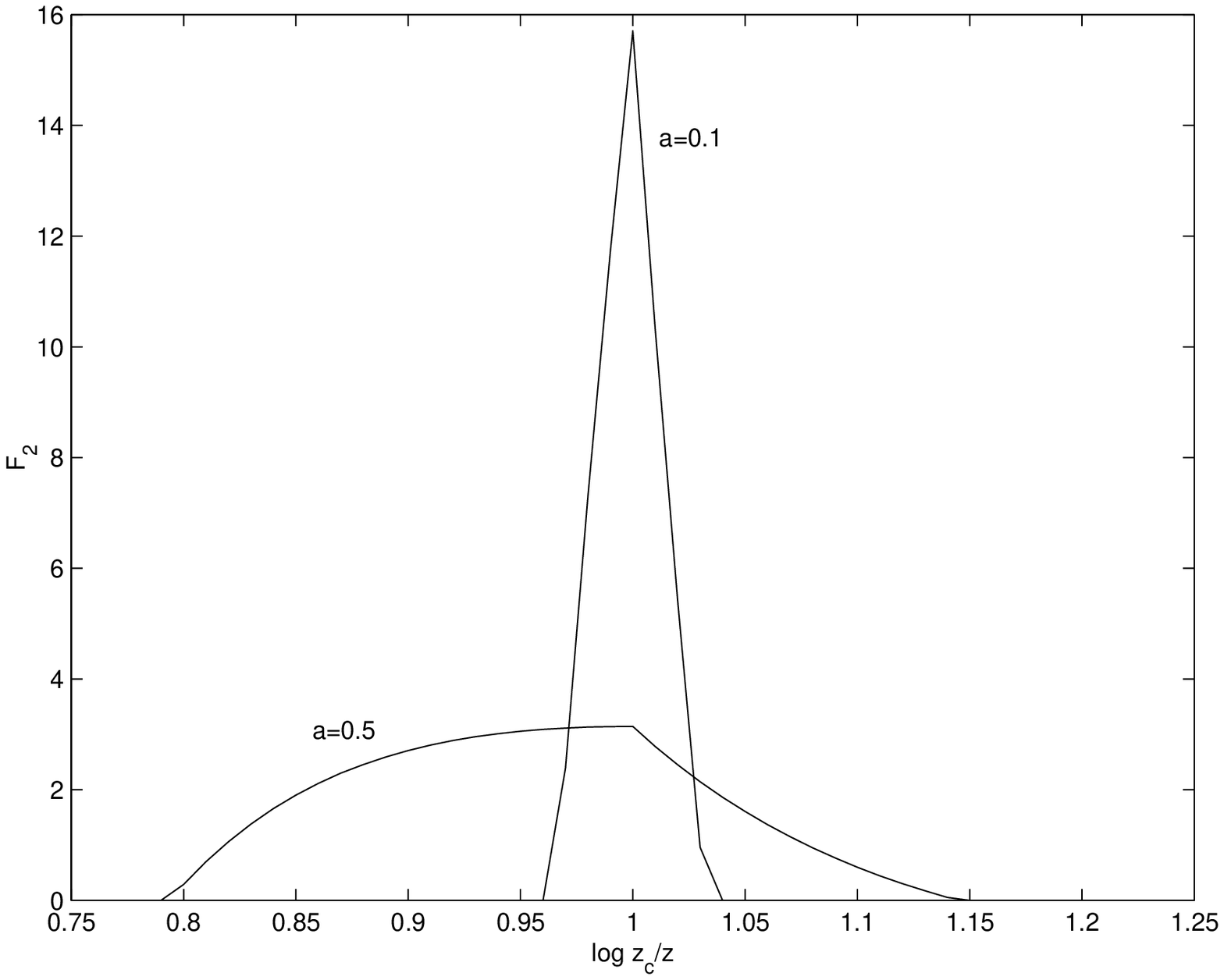}}
\caption[]{$F_1$ and $F_2$ as functions of distance along the wave
trajectory given the particle distribution function in the form
(32) }
\end{figure*}
\clearpage
\begin{figure*}
\input epsf
\def\epsfsize#1#2{0.35#1} \centerline{\epsfbox{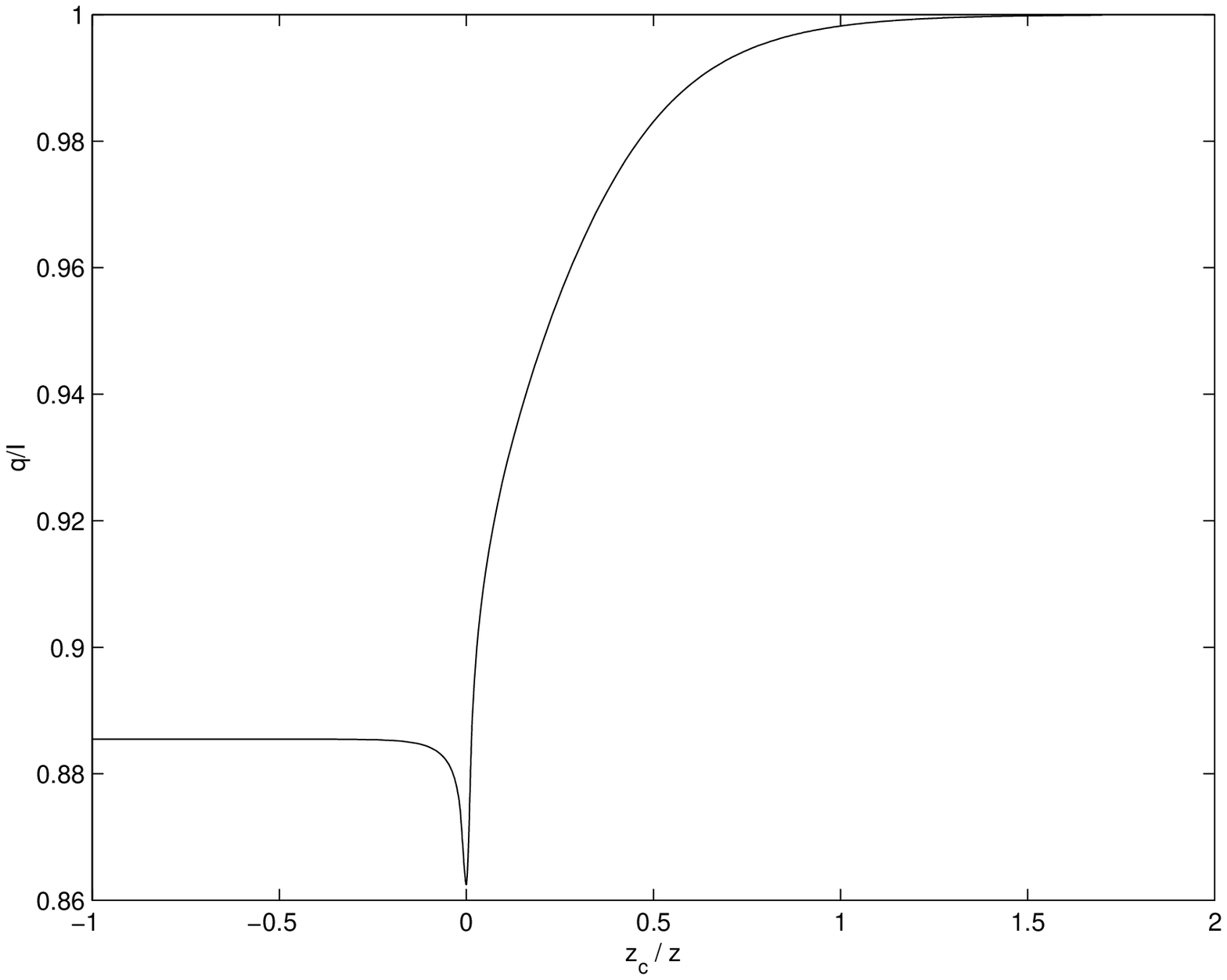}}
\input epsf
\def\epsfsize#1#2{0.35#1} \centerline{\epsfbox{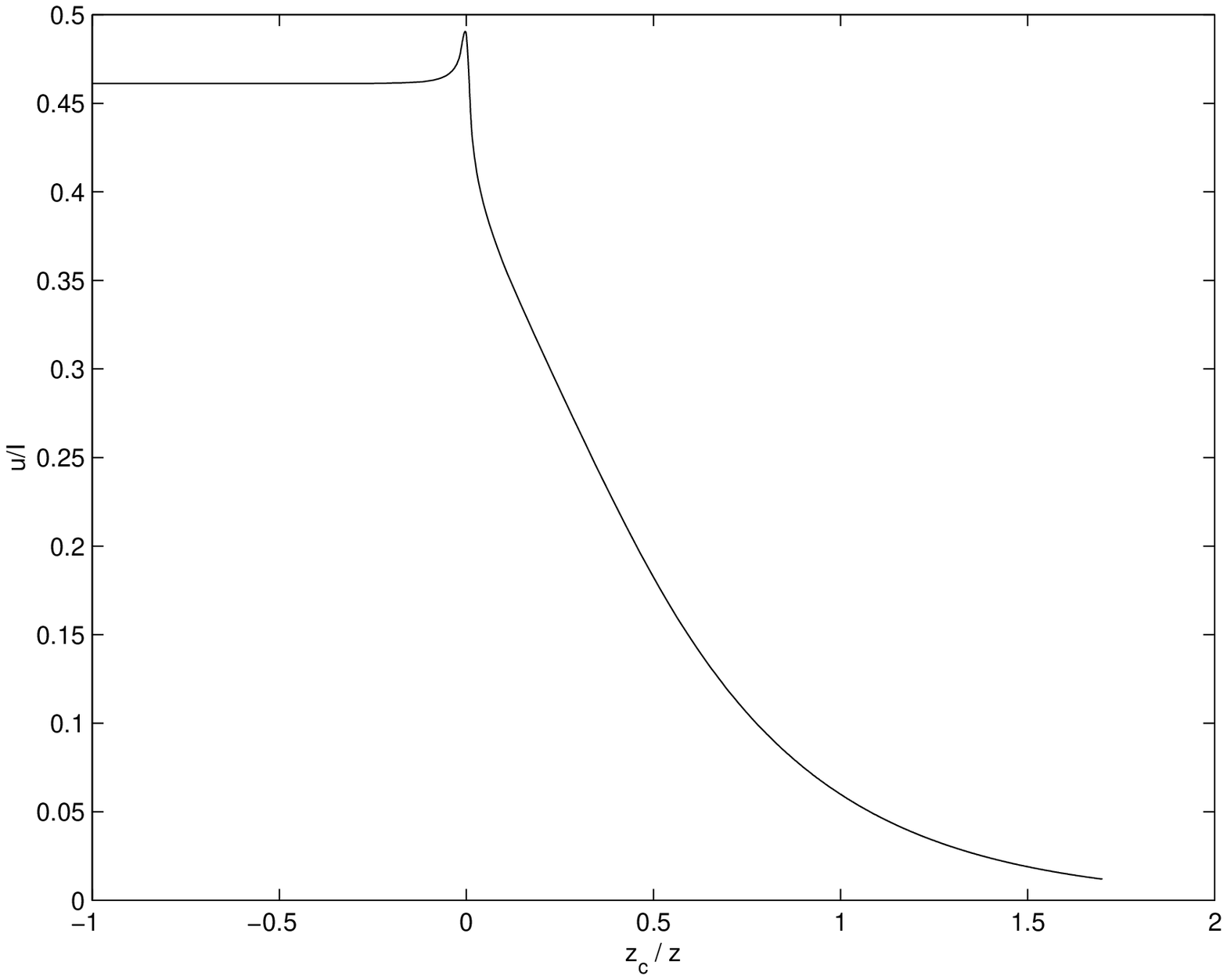}}
\input epsf
\def\epsfsize#1#2{0.35#1} \centerline{\epsfbox{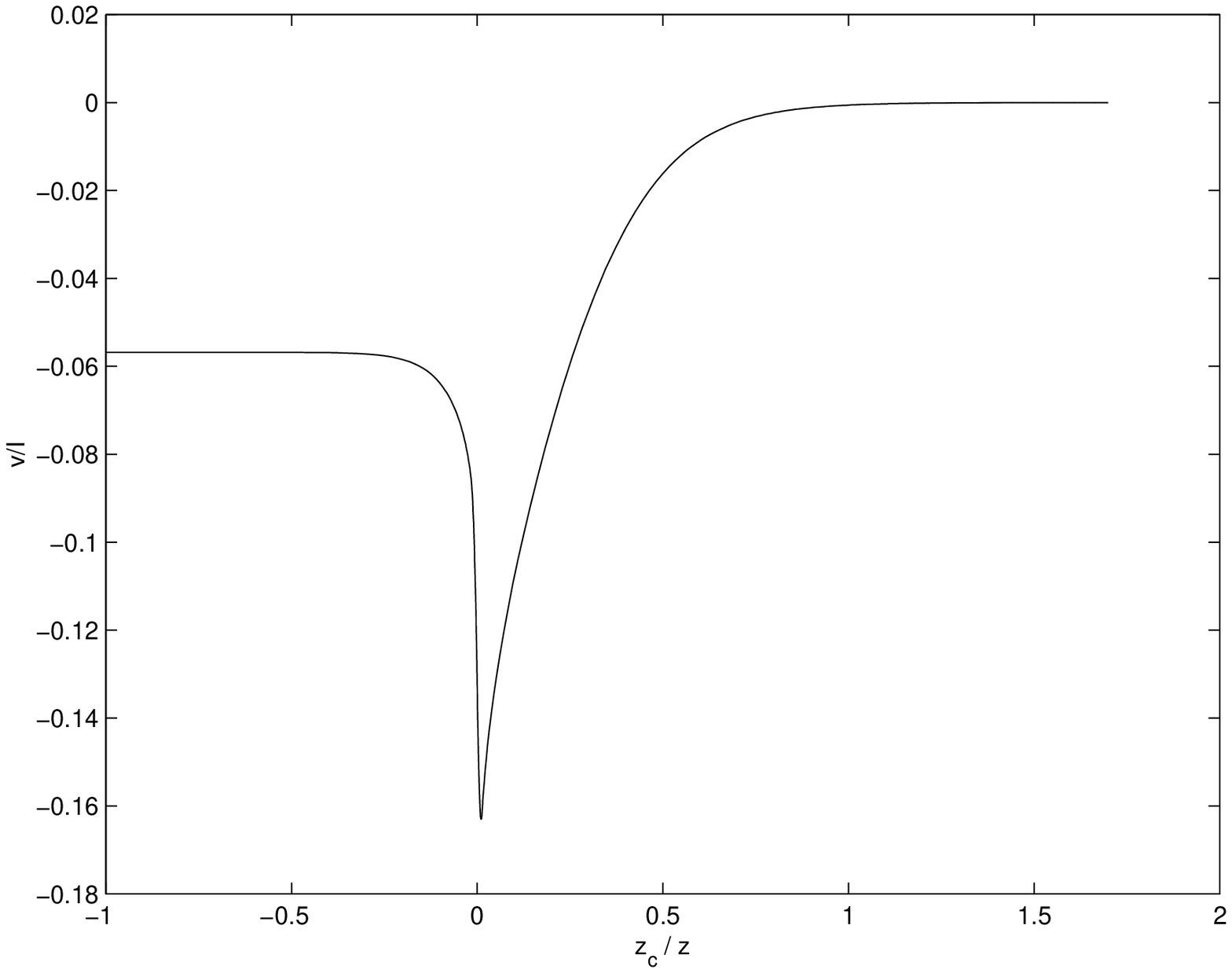}}
\end{figure*}
\clearpage
\begin{figure*}
\input epsf
\def\epsfsize#1#2{0.35#1} \centerline{\epsfbox{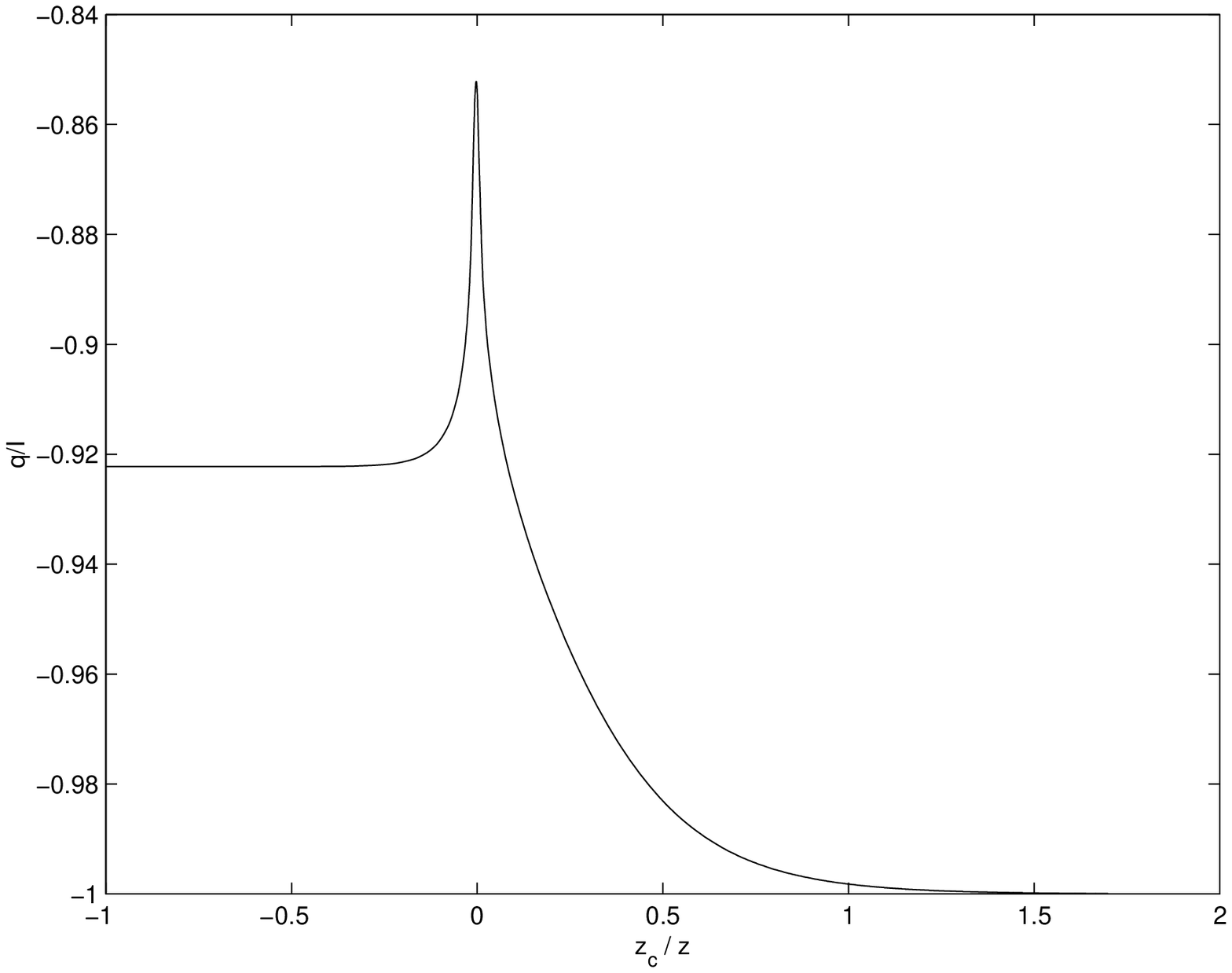}}
\input epsf
\def\epsfsize#1#2{0.35#1} \centerline{\epsfbox{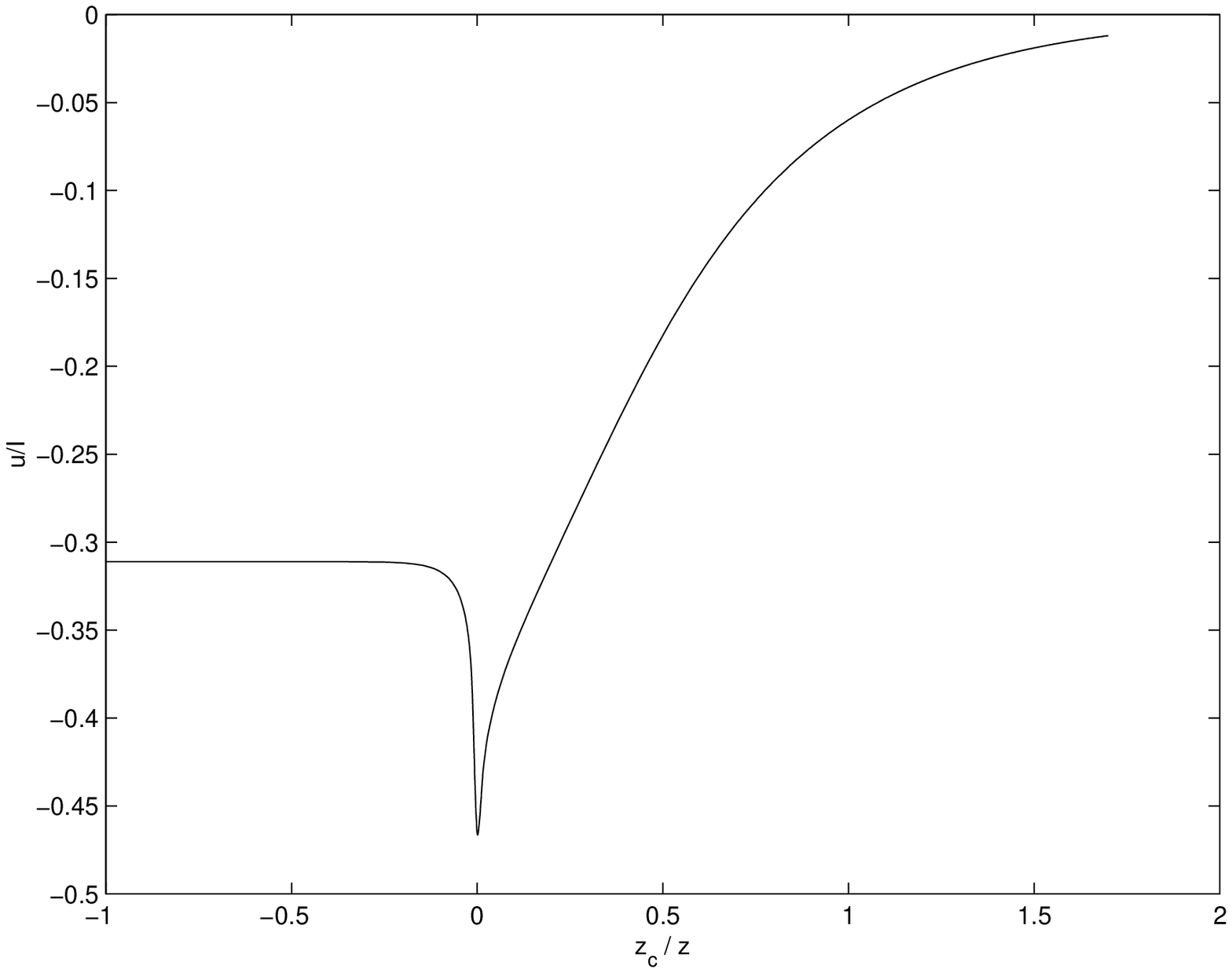}}
\input epsf
\def\epsfsize#1#2{0.35#1} \centerline{\epsfbox{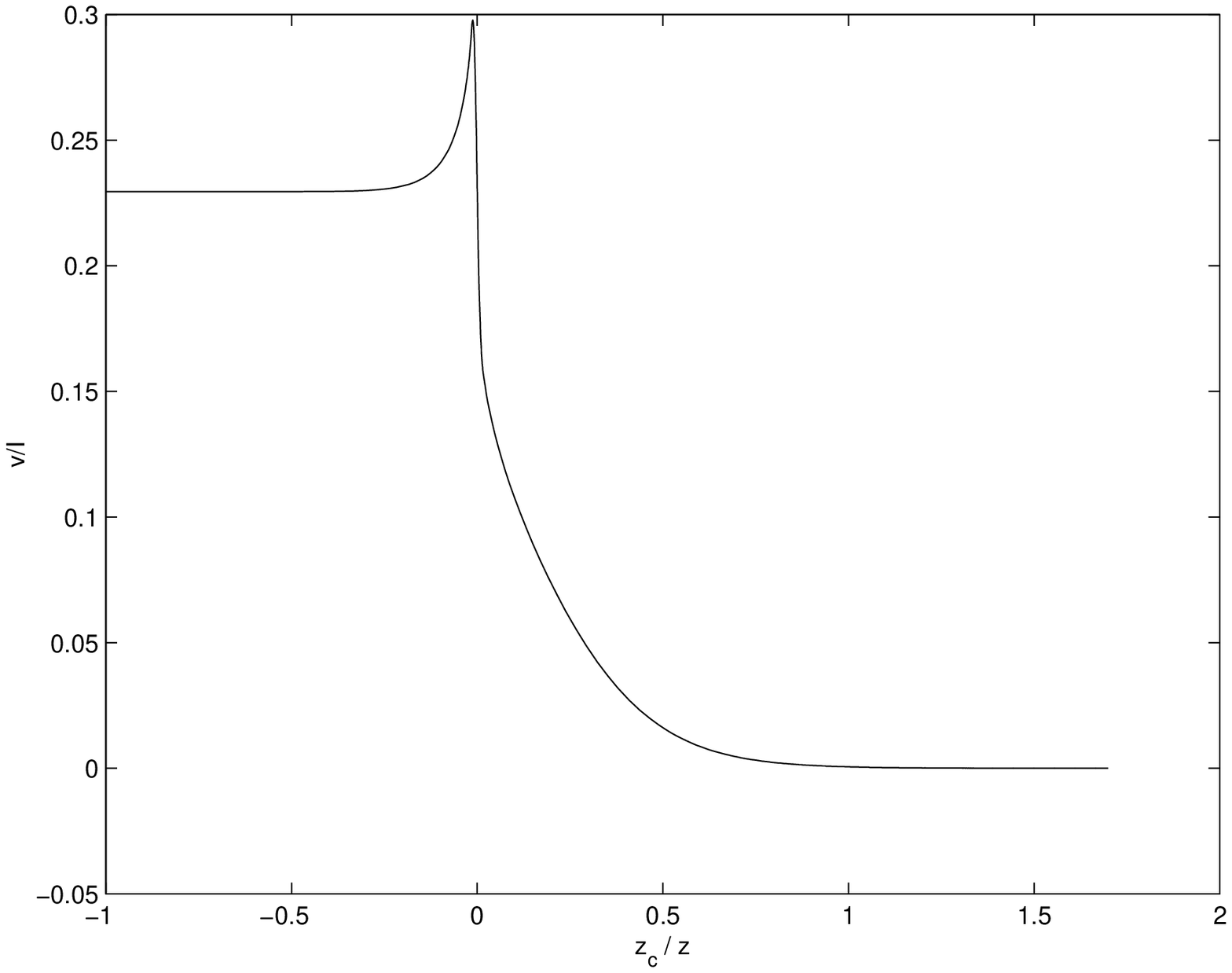}}
\caption[]{Evolution of the normalized Stokes parameters of the
original ordinary ({\bf a-c}) and extraordinary ({\bf d-f}) waves
along the trajectory; $\eta =1$, $\mu =0.3$, $a=0.1$.}
\end{figure*}

\clearpage

\begin{figure*}
\input epsf
\def\epsfsize#1#2{0.5#1} \centerline{\epsfbox{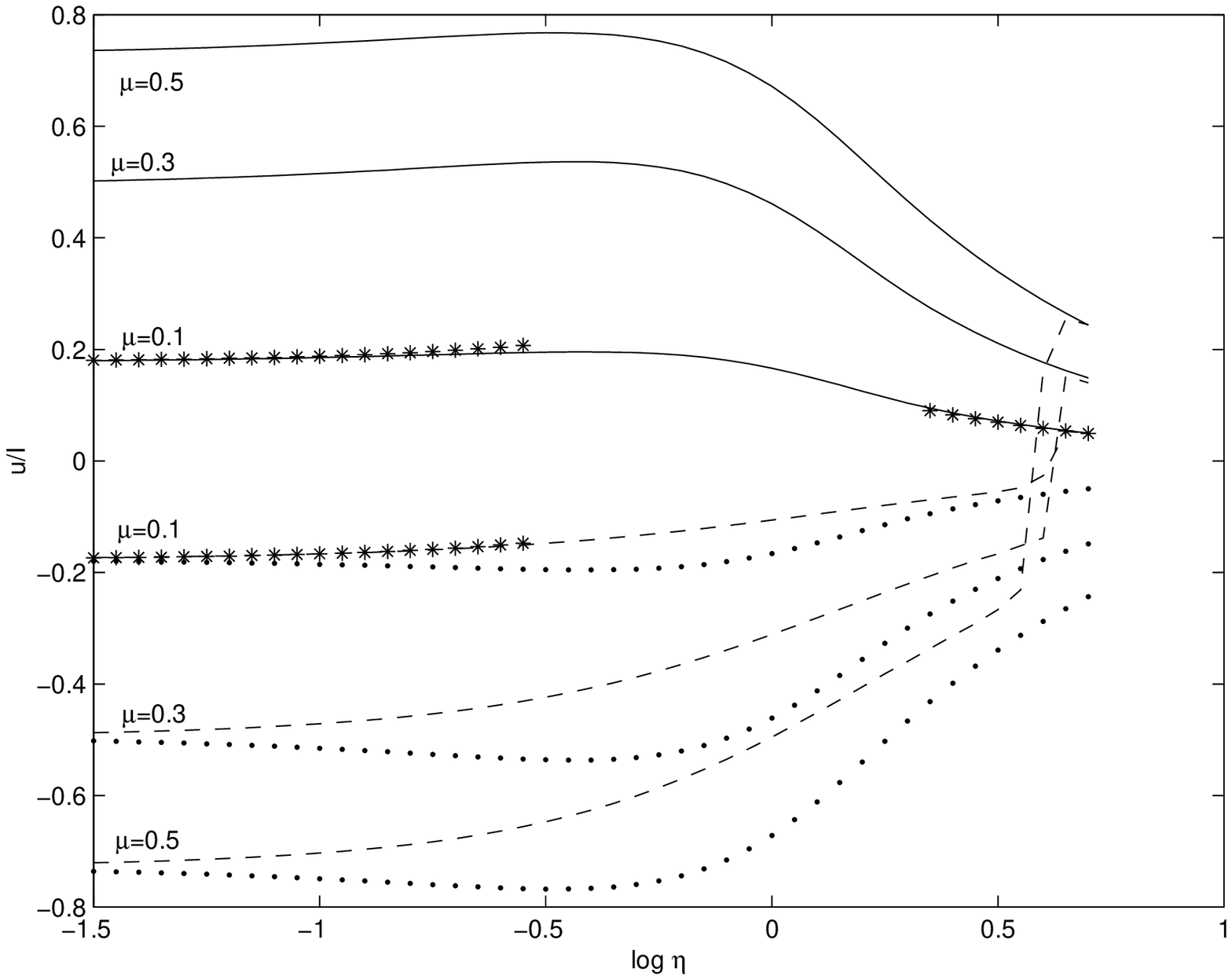}}
\input epsf
\def\epsfsize#1#2{0.5#1} \centerline{\epsfbox{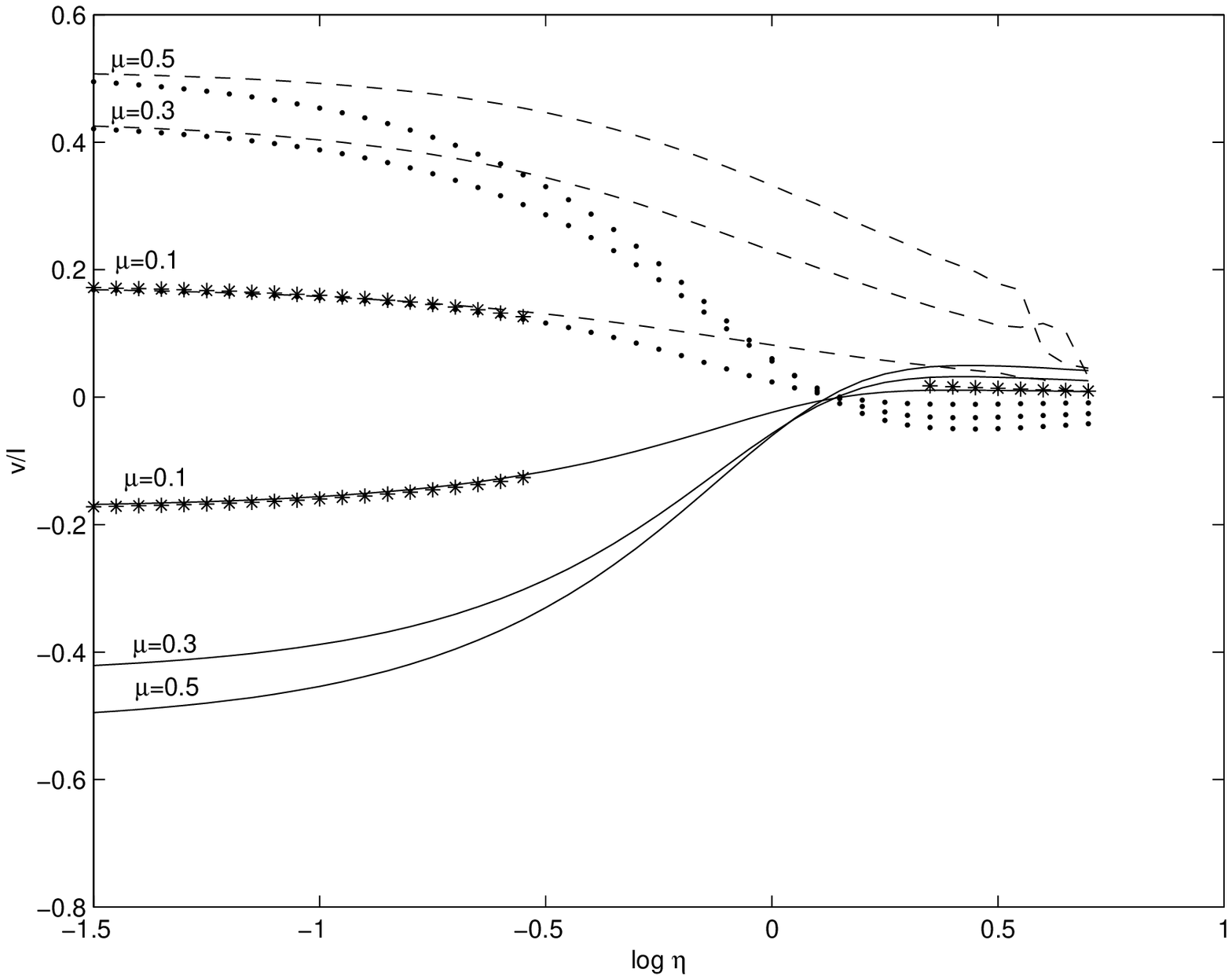}}
\caption[]{The final normalized Stokes parameters as functions of
$\eta$ for different $\mu$. The analytical approximations given by
equations (30) and (31) are shown in asterisks. The solid and
dashed lines correspond to the original ordinary and extraordinary
waves, respectively. The dotted lines show the parameters
orthogonal to those of the ordinary mode. }
\end{figure*}

\clearpage

\appendix

\section{Conductivity of the hot magnetized plasma}
Our aim here is to find the conductivity tensor of the hot
magnetized plasma in the coordinate system with the z-axis along
the wavevector. In the commonly used system with the z-axis along
the magnetic field direction, it has the following form
\citep[e.g. ][]{Mikh75}:
\begin{equation}\hat\sigma_0=-ie^2\times\left <\sum_{n=-\infty}^\infty\zeta_n\Phi_\perp\left (
\begin{array}{lll}v_\perp^2\frac{\displaystyle n^2}{\displaystyle \xi^2}J_n^2 &
iv_\perp^2\frac{\displaystyle nJ_nJ_n^\prime}{\displaystyle \xi} &
v_\perp v_z\frac{\displaystyle nJ_n^2}{\displaystyle \xi}\\
-iv_\perp^2\frac{\displaystyle nJ_nJ_n^\prime}{\displaystyle \xi}
& v_\perp^2J_n^{\prime 2} & -iv_\perp v_zJ_nJ_n^\prime \\v_\perp
v_z\frac{\displaystyle nJ_n^2}{\displaystyle \xi} & iv_\perp
v_zJ_nJ_n^\prime & v_zJ_n^2\Phi_\Vert /\Phi_\perp
  \end{array}
  \right )\right >\,.
  \end{equation}
  Here the brackets $<...>$ stand for the integral operator
  $\int ...p_\perp{\rm d}p_\perp{\rm d}p_z$; $J_n$ and
  $J_n^\prime$ are  the Bessel function of the $n$-th order and its
  derivative, $\xi$ is their argument, \[\xi=k_\perp v_\perp\gamma
  /\omega_H\,;\] \[\zeta_n=(\omega
  -k_zv_z-n\omega_H/\gamma)^{-1}\,;\]
  \[\Phi_\perp=\frac{1}{v_\perp}\frac{\partial F}{\partial p_\perp}
  +\frac{k_z}{\omega}\left (\frac{\partial F}{\partial p_z}-
  \frac{v_z}{v_\perp}\frac{\partial F}{\partial p_\perp}\right
  )\,;\]
  \[\Phi_\Vert =\frac{\partial F}{\partial p_z}-\frac{n\omega_H}
  {\gamma\omega}\left (\frac{\partial F}{\partial p_z}-
  \frac{v_z}{v_\perp}\frac{\partial F}{\partial p_\perp}\right
  )\,;\]
  $F$ is the particle distribution function with the normalization
  \[\int F(p_\perp\,,p_z)p_\perp{\rm d}p_\perp{\rm d}p_z=N_0\,,\] and
  the subscripts "z" and "$\perp$" refer to the components
  parallel and perpendicular to the magnetic field, respectively.

  In terms of the quantities in the system with $Oz\Vert {\bf k}$,
  $v_z=v_\Vert$, $k_z=kb_z$, $k_\perp =kb_\perp$, $b_\perp
  =\sqrt{b_x^2+b_y^2}$. Then the argument of the Bessel function
  is written as \[\xi=\frac{kb_\perp v_\perp\gamma}{\omega_H}=
  \frac{\omega^\prime}{\omega_H}\frac{\psi}{b_\perp}\,,\] where
  ${\omega^\prime =\omega\gamma (b_x^2+b_y^2)}$ is the frequency
  in the rest frame of the particle, $\psi$ the particle
  pitch-angle and it is taken into account that $kv/\omega\approx
  1$, i.e. $v\approx c$ and the refractive index is approximately
  unity. Hereafter it is assumed that the particle pitch-angle is
  small, $\psi\ll b_\perp$, in which case $\xi\ll 1$ up to the
  cyclotron resonance, $\omega^\prime =\omega_H$. As can be seen
  from the final equations (6), beyond the cyclotron resonance, at
  $\omega^\prime >\omega_H$, the evolution of the electric field
  amplitudes rapidly ceases because of strong decrease of
  $\omega_H$, $\omega_H\propto r^{-3}$. Therefore in the case of
  interest $\xi\ll 1$ and one can use the approximation of the
  Bessel function: \[J_n\approx\frac{(\xi /2)^n}{n!}\,.\] Then the
  form of the conductivity tensor is substantially simplified:
  \[\sigma_{0_{11}}=\sigma_{0_{22}}=-ie^2\left <
  \frac{\gamma\omega(1-\beta b_z)^2}{m\left (\omega_H^2-
  \omega^{\prime 2}\right )}\right >\,,\]
\[\sigma_{0_{12}}=-\sigma_{0_{21}}=e^2\left <
  \frac{\omega_H(1-\beta b_z)}{m\left (\omega_H^2-
  \omega^{\prime 2}\right )}\right >\,,\]
\begin{equation}
\sigma_{0_{13}}=\sigma_{0_{31}}=-ie^2\left <
  \frac{\gamma\omega b_\perp\beta(1-\beta b_z)}{m\left (\omega_H^2-
  \omega^{\prime 2}\right )}\right >\,,
  \end{equation}
\[\sigma_{0_{23}}=\sigma_{0_{32}}=-e^2\left <
  \frac{\omega_Hb_\perp\beta}{m\left (\omega_H^2-
  \omega^{\prime 2}\right )}\right >\,,\]
  \[\sigma_{0_{33}}=-ie^2\left <
  \frac{\gamma\omega b_\perp^2\beta^2}{m\left (\omega_H^2-
  \omega^{\prime 2}\right )}-\frac{\omega}{m\gamma
  \omega^{\prime 2}}\right >\,,\]
  where the brackets stand for the integral operator $\int ...F(p_\Vert){\rm
  d}p_\Vert$.

  In the system with $Oz\Vert {\bf k}$, the conductivity tensor is
  given by
  \begin{equation}
  \hat\sigma =\hat R\hat\sigma_0\hat R^{-1}\,,
  \end{equation}
  where $\hat R$ and $\hat R^{-1}$ are the matrices of the direct
  and inverse transformation between the coordinate systems
  considered:
  \begin{equation}
  \hat R=\left (\begin{array}{ccc}
  -\frac{\displaystyle b_xb_z}{\displaystyle b_\perp}
   & \frac{\displaystyle b_y}{\displaystyle b_\perp} & b_x\\
-\frac{\displaystyle b_yb_z}{\displaystyle b_\perp} &
-\frac{\displaystyle b_x}{\displaystyle b_\perp} & b_y\\ b_\perp &
0 & b_z
  \end{array}\right )\,,
  \end{equation}
  \begin{equation}
  \hat R^{-1}=\left (\begin{array}{ccc}
  -\frac{\displaystyle b_xb_z}{\displaystyle b_\perp
  } & -\frac{\displaystyle b_yb_z}{\displaystyle b_\perp} & b_\perp\\
  \frac{\displaystyle b_y}{\displaystyle b_\perp}
  & -\frac{\displaystyle b_x}{\displaystyle b_\perp} & 0\\
  b_x & b_y & b_z
  \end{array}\right )\,.
  \end{equation}
  As is discussed after equation (5), the waves are almost transverse
  and we take $E_z=0$. Then we should find only
  \begin{equation}
  j_x=\sigma_{xx}E_x+\sigma_{xy}E_y\quad {\rm and}\quad
  j_y=\sigma_{yx}E_x+\sigma_{yy}E_y\,.
  \end{equation}
  Substituting equations (A4) and (A5) in equation (A3) and taking into
  account that $\sigma_{0_{11}}=\sigma_{0_{22}}$,
  $\sigma_{0_{12}}=-\sigma_{0_{21}}$, $\sigma_{0_{13}}=\sigma_{0_{31}}$
  and $\sigma_{0_{23}}=-\sigma_{0_{32}}$, one obtains:
  \[\sigma_{xx}=\left (1-b_x^2\right )
  \sigma_{0_{11}}+b_x^2\sigma_{0_{33}}-\frac{2b_x^2b_z}{b_\perp}\sigma_{0_{13}}\,,\]
\[\sigma_{yy}=\left (1-b_y^2\right )
  \sigma_{0_{11}}+b_y^2\sigma_{0_{33}}-\frac{2b_y^2b_z}{b_\perp}\sigma_{0_{13}}\,,\]
  \[\sigma_{xy}=b_xb_y\left (-\sigma_{0_{11}}+\sigma_{0_{33}}-
  2(b_z/b_\perp)\sigma_{0_{13}}\right
  )+b_z\sigma_{0_{12}}+b_\perp\sigma_{0_{23}}\,,\]
\begin{equation}
\sigma_{yx}=b_xb_y\left (-\sigma_{0_{11}}+\sigma_{0_{33}}-
  2(b_z/b_\perp)\sigma_{0_{13}}\right )-b_z\sigma_{0_{12}}
  -b_\perp\sigma_{0_{23}}\,,
  \end{equation}
  Using equation (A2) in equation (A7) yields finally:
  \[\sigma_{xx}=a_1b_x^2-a_2\,,\quad \sigma_{yy}=a_1b_y^2-a_2\,,\]
\begin{equation}
\sigma_{xy}=a_1b_xb_y+g\,,\quad
\sigma_{yx}=a_1b_xb_y-g\,,
\end{equation} with \[a_1=\left
<\frac{ie^2\omega\omega_H^2}{m\omega^{\prime 2}\gamma\left
(\omega_H^2-\omega^{\prime 2}\right ) }\right
>\,,\]
\[a_2=\left <\frac{-ie^2\gamma\omega (1-\beta b_z)^2)}{m\left
(\omega_H^2-\omega^{\prime 2}\right ) }\right
>\,,\]
\[g=\left
<\frac{e^2\omega_H(b_z-\beta)}{m\left(\omega_H^2-\omega^{\prime
2}\right ) }\right >\,.\] Throughout this section, we have
considered the one-component plasma for simplicity. The results
can be extended to the case of pair plasma by introducing the
summation over the particle species.


\end{document}